\documentclass[9pt, a4paper, twocolumn]{extarticle}
\usepackage{graphicx}
\usepackage{amsmath, amsfonts}
\usepackage{graphicx}
\usepackage{hyperref}
\usepackage[a4paper, total={7in, 9in}]{geometry}
\usepackage[affil-it]{authblk}
\usepackage{tikz, standalone}
\usepackage[numbers, sort&compress,merge]{natbib}
\usepackage{svg}
\usepackage{xcolor}
\usepackage{caption}
\usepackage{subcaption}
\usetikzlibrary{calc}
\usetikzlibrary{backgrounds, decorations.pathreplacing}
\title{Persistent pseudopod splitting is an effective chemotaxis strategy in shallow gradients}

\author[1]{Albert Alonso}
\author[1,2]{Julius B. Kirkegaard}
\author[3]{Robert G. Endres \thanks{Corresponding author: r.endres@imperial.ac.uk}}
\affil[1]{Niels Bohr Institute, University of Copenhagen, Denmark}
\affil[2]{Department of Computer Science, University of Copenhagen, Denmark}
\affil[3]{Department of Life Sciences and Centre for Integrative Systems Biology and Bioinformatics, Imperial College, London, United Kingdom}

\begin{document}
\twocolumn[\begin{@twocolumnfalse}

\maketitle
\begin{abstract}
Single-cell organisms and various cell types use a range of motility modes when following a chemical gradient, but it is unclear which mode is best suited for different gradients.
Here, we model directional decision-making in chemotactic amoeboid cells as a stimulus-dependent actin recruitment contest.
Pseudopods extending from the cell body compete for a finite actin pool to push the cell in their direction until one pseudopod wins and determines the direction of movement.
Our minimal model provides a quantitative understanding of the strategies cells use to reach the physical limit of accurate chemotaxis, aligning with data without explicit gradient sensing or cellular memory for persistence.
To generalize our model, we employ reinforcement learning optimization to study the effect of pseudopod suppression, a simple but effective cellular algorithm by which cells can suppress possible directions of movement.
Different pseudopod-based chemotaxis strategies emerge naturally depending on the environment and its dynamics.
For instance, in static gradients, cells can react faster at the cost of pseudopod accuracy, which is particularly useful in noisy, shallow gradients where it paradoxically increases chemotactic accuracy.
In contrast, in dynamics gradients, cells form \textit{de novo} pseudopods. 
Overall, our work demonstrates mechanical intelligence for high chemotaxis performance with minimal cellular regulation.
\end{abstract}

\thanks*{\footnotesize Corresponding author: r.endres@imperial.ac.uk}
\vspace*{0.8cm}\end{@twocolumnfalse}]

\section{Introduction}
Mechanical intelligence is widespread in nature, by which information processing is deeply embedded in the architecture of living systems \cite{wan_origins_2021, bodor_cell_2020}.
For instance, the underlying mechanisms by which cells perform chemotaxis, the directed movement of an organism along a chemical concentration gradient during microbial pathogenesis, wound healing, and immune response, remains a subject of intensive research \cite{grognot_physiological_2023, alonso_learning_2024, brumley_bacteria_2019, varennes_collective_2016}.
Particularly relevant is the understanding of the tight coupling between sensory cues and cell locomotion mechanisms, as they provide insights into effective navigation methods at the microscopic scale for cell sensing at fundamental physical limits \cite{wan_active_2023}.
Here, we focus on studying the role of pseudopod formation as a cellular decision-making mechanism, which represents an important yet not fully understood aspect of cellular navigation \cite{tweedy_distinct_2013, andrew_chemotaxis_2007}.

Amoeboid locomotion is characterized by the extension of pseudopods, temporary protrusions that allow the cell to explore its environment and move directionally in response to chemical cues \cite{wan_origins_2021}.
Experimental evidence has shown pseudopod formation to occur at higher rates in shallow gradients and that they are more pronounced \cite{van_haastert_chemotaxis_2010}.
In particular, this phenomenon may constitute a way to reach the fundamental physical limit by minimizing the interference of sensing by movement \cite{tweedy_distinct_2013}. 
The ultimate limit is reached by a cell that senses only previously undetected ligand molecules to gain new information, effectively corresponding to a ligand-absorbing cell \cite{tweedy_distinct_2013}.
Despite the observed importance of amoeboid cell migration, a comprehensive theoretical understanding of pseudopod splitting and its strategic role in accurate chemotaxis is still lacking. Traditionally, chemotaxis is approached from microscopic receptor-ligand interactions at the cell surface, translating external signals into directional movement through a complex network of signaling pathways \cite{bosgraaf_navigation_2009, swaney_eukaryotic_2010}.
Often, models assume an "all-knowing" cell, capable of optimally processing their signaling information for directional migration\cite{alonso_learning_2024, endres_accuracy_2008, rode_information_2024}.
Alternative, more realistic models exist with a tight coupling between signaling, cytoskeleton remodeling, and cell-shape dynamics \cite{tweedy_distinct_2013, mogilner_intracellular_2018}, but from which insights are more difficult to obtain.

Here, we model pseudopod splitting through the dynamics of actin polymerization, wherein the competition for a finite resource between extending pseudopods determines the next cell movement direction.
By quantitatively describing intracellular interactions within an interpretable model, we gain insights into the fundamental principles governing cellular decision-making and their implications for efficient chemotaxis in complex chemical landscapes.
To generalize, we employ state-of-the-art deep reinforcement learning (DRL) \cite{sutton_reinforcement_2018, schulman_proximal_2017}, allowing us to study how pseudopod suppression enhances the cell's ability to correctly choose the direction of movement faster.
By optimizing a self-contained but unconstrained suppression policy at a level beyond what is possible with classical optimization, we achieve chemotactic strategies that cells have evolved over evolutionary time scales.
One example is the experimentally observed alternation of cells between pseudopod splitting and elongation \cite{tweedy_distinct_2013}.
Deciphering the key physical principles of embodied computation in the living world allows us to understand the cell body as an analog machine for both information processing and motility.
This approach may inspire entirely new classes of intelligent matter designs.

\section{Decision-making model}
\begin{figure*}[tbh]
    \centering
     \includegraphics[width=13cm]{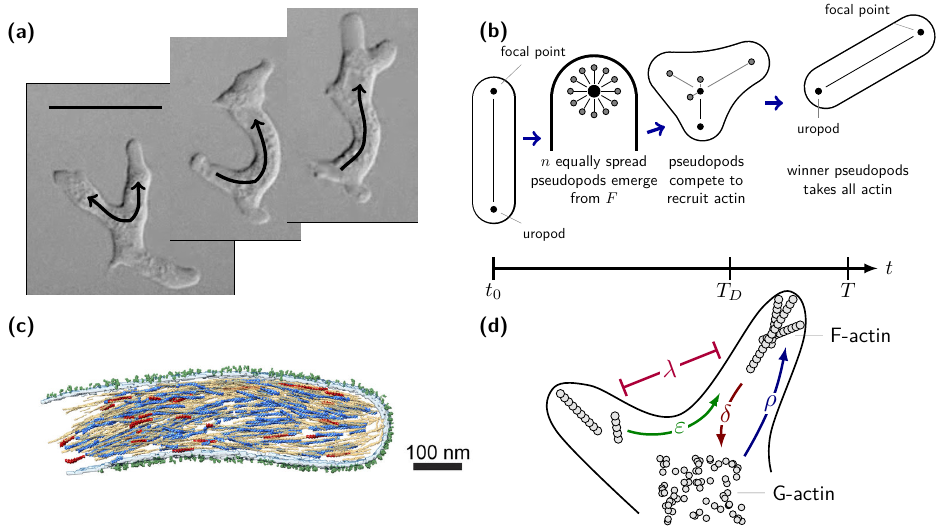}
     \caption{\textbf{Schematics of the decision-making model and actin dynamics.}
     \textbf{(a)}~Pseudopod splitting in a \textit{Dictyostelium} cell performing chemotaxis under agar (provided by Robert Insall) \cite{andrew_chemotaxis_2007}.
     Arrows indicate likely actin flows, and the scale bar is 20\textmu m.
     \textbf{(b)}~Diagram of the cell morphology during a \textsl{splitting event}.
     Pseudopod formation occurs due to competition during actin recruitment.
     In this instance, only two pseudopods emerge even though $n=12$ candidates start the competition.
     Finally, only one remains after decision-making time $T_D$, altering the cell orientation and advancing its position.
     The movement step is completed at time $T$.
     \textbf{(c)}~Segmentation of actin filaments inside pseudopodia of small platelets with permission from \citet{sorrentino_structural_2021} based on high-resolution structural analysis.
     Actin filaments are in blue, red, and yellow, while receptors are shown in green.
     \textbf{(d)}~Schematics of the simplified polymerization of actin into filaments (F-actin) at the internal membrane surface from actin monomers (G-actin), forming pseudopods. 
     The diagram also shows how pseudopods suppress neighbors by redistributing their actin filaments and mutually inhibiting each other's growth.
     }
     \label{fig:diagram-splitting-event}
\end{figure*}

In the context of cellular decision-making, pseudopods play an important role that extends far beyond mere movement \cite{tweedy_distinct_2013} -- we consider them fundamentally coupled with the sensing process.
Experimental results show pseudopod formation originates mainly from two distinct mechanisms: splitting from existing pseudopods or by \textit{de novo} formation (Fig.~\ref{fig:diagram-splitting-event}a) \cite{van_haastert_chemotaxis_2010}.
To simplify this complexity, our minimal model assumes a unified cell body and pseudopod, where the previous pseudopod serves as the origin for new pseudopod growths, which permits us to describe both mechanisms under the same dynamical framework.
We refer to this as a \textsl{splitting event} (Fig.~\ref{fig:diagram-splitting-event}b).
A splitting event is defined as a competitive process between $n$ possible directions for the cell to move (or grow, in our case), where the winning candidate dictates the new cell orientation.
We allow 12 directional options, providing a suitable number of choices for the cell to navigate its environment.
In order to win, each candidate attempts to polymerize as many actin filaments as possible from a finite reservoir of actin monomers.

\paragraph{Actin polymerization.}
Previous work by \citet{pais_mechanism_2013} successfully modeled collective decision-making on honeybee swarms using stochastic differential equations for valued-based decisions with a finite resource to distribute, e.g., swarm members to 2 or more potential nest sites.
Similarly, our model describes actin polymerization on each pseudopod (Fig.~\ref{fig:diagram-splitting-event}c) with the following overdamped Langevin dynamics
\begin{align}\label{eq:actin-dynamics}
    \dfrac{dA_i}{dt} &= \rho_i A_u - \delta A_i - \lambda A_i \bar{A_i} + \varepsilon (A_i - \bar{A_i}) + \eta_i(t),
\end{align}
where $A_i$ are the cumulative proportion of actin monomers that have been polymerized into filaments at pseudopod $\mathcal{P}_i$, and $\bar{A_i}$ are the local sum of actin levels of the other pseudopods $\bar{A_i} = \sum_{j\ne i} A_j$.
Assuming actin mass conservation, we set $A_u = 1 - \sum_{i} A_i$ as the uncommitted actin monomers (G-Actin) proportion inside the cell.

Fluctuations are assumed to arise by the Berg-Purcell noise in chemoattractant measurements \cite{berg_physics_1977}, which we include by adding noise $\eta$ with 
\begin{equation}\label{eq:actin-noise}
    \langle \eta_i(t) \rangle = 0, \qquad \langle\eta_i(t)\cdot \eta_j(t^\prime)\rangle~=~\sigma\cdot c_i\cdot \delta(t-t^\prime) \cdot \delta_{ij},
\end{equation}
where $\delta(\cdot)$ is Dirac's delta, $\delta_{ij}$ is Kronecker's delta, and $c_i \equiv c(x_i)$ is the dimensionless concentration level at the tip of pseudopod $\mathcal{P}_i$ (see Methods for details on non-dimensionalization).
This assumes ligand noise as Poissonian due to the random arrival of ligand molecules at the cell surface by diffusion, and internal noise due to fluctuations in actin levels is ignored.

The polymerization rate $\rho_i$ of actin filaments inside a given pseudopod $\mathcal{P}_i$ is influenced by the local concentration of chemoattractant, which activates well-known signaling pathways that lead to increased actin polymerization \cite{van_haastert_chemotaxis_2010}.
Thus, we model the amount of G-actin recruited by a pseudopod as being proportional to the signaling activity of the receptors at the pseudopod end.
The rate is defined as the average signaling activity of a receptor complex composed of many receptors, which can switch between an on and an off state \cite{clausznitzer_predicting_2014}, given by the Boltzmann probability $P_{on}$ with a strength $\rho_0$
\begin{equation}\label{eq:polymerization-rate}
    \rho_i = \rho_0 \cdot P_{on} = \dfrac{\rho_0}{1 + e^{\Delta F_i}},
\end{equation}
where we linearized the change in free energy for small ligand concentration changes, i.e., $\Delta F_i\approx-\kappa_c \left(c_i - c_0 \right)$, with $c_i$ and $c_0$ being the concentration value at the end of the pseudopod and at the original focal adhesion point, respectively.
Note that in an environment without chemoattractant gradient, i.e., constant concentration profile, all pseudopods have the same intrinsic polymerization rate set by $\rho = \rho_0 \mathbin{/} 2$.

F-actin constantly undergoes treadmilling, where individual monomers are removed from one end to be added at the other end of the polymer (Fig.~\ref{fig:diagram-splitting-event}d).
We include this by adding a depolymerization rate ($\delta$).
In our case, however, once the monomer has left the filament, we consider it to be returning to the uncommitted actin pool and, thus, potentially being reused by other pseudopods.
Furthermore, pseudopods may inhibit each other by sequestering shared resources and signaling crosstalk \cite{davidson_wasp_2017}.
Hence, we also include a cross-inhibition ($\lambda$) term, where the size of the rival candidates will diminish the overall recruitment speed \cite{pais_mechanism_2013}.

Finally, the actin exchange rate ($\varepsilon$) represents the transfer of actin between pseudopods (Fig.~\ref{fig:diagram-splitting-event}d), as it has been observed that cells can redistribute actin to prioritize certain directions \cite{ecker_excitable_2021, yolland_persistent_2019}.
This term results in a \textsl{commitment to the winner} behavior where the cell follows the largest pseudopod the moment it has grown enough to collapse the other candidates back to the focal adhesion point.

\paragraph{Pseudopod growth.}
To model pseudopod growth, we assume actin to be the sole driver of membrane expansion, thereby linking sensing mechanisms to cellular motility.
Hence, by focusing on its intrinsic coupling, we ignore some known effects of membrane mechanics on cellular motility, such as membrane tension, substrate interaction, or surface curvature.
Due to the high fluctuations in the actin dynamics at elevated chemoattractant concentrations, we model pseudopod length as a timed-average linear response of F-actin levels, such as 
\begin{equation}\label{eq:length-linear-response}
    \ell_i(t)~=~L \int_{t-1}^{t}  A_i(t') dt^\prime,
\end{equation}
such that the total dimensionless length of the cell is conserved, i.e., $L~=~\ell_u{+}\sum_i^n \ell_i$, with the time for the linear filter used as the characteristic timescale for the dynamics (see Methods for further details).

\paragraph{Decision time.}
At the beginning of a \textsl{splitting event}, we consider all the motility-associated actin to be unpolymerized $A_i(t_0) = 0$ for all $i \in \{1,\ldots,n\}$, and $A_u(t_0) = 1$.
The event ends at $t=T$ when one candidate has gathered most of the actin ($A_i \approx 0.95$).
However, the duration of the event contains both the decision-making process and the final growth of the winning pseudopod until it takes all remaining actin.
Since we are interested in the decision time $T_D$, we define it as the time beyond which the length of the winning pseudopod is larger than the summed lengths of all remaining candidate pseudopods, i.e.
\begin{equation}\label{eq:decision_time}
    \ell_i(t) \ge \sum_{j\neq i}\ell_j(t), \quad \forall t \ge T_D.
\end{equation}

\paragraph{Concentration profile.}
To simulate chemotactic environments, we assume a linear gradient of concentration profile similar to those observed in chemotactic chambers \cite{tweedy_distinct_2013}, such as
\begin{equation}\label{eq:concentration-profile}
    c(x) = g_x x + c_n.
\end{equation}
where $g_x$ is the concentration gradient and $c_n$ is the background concentration at the origin.
This is a common approximation to the resulting concentration profile based on Flick's second law of diffusion of a constant chemoattractant value that diffuses from the side of the chamber.
We treat gradient magnitude and background concentration as freely varying initial conditions to investigate their impact on cell decision-making processes.
Notably, the profile is defined with unitless variables as described in the Methods section.

\section{Results}

\begin{figure}[tbh]
    \centering
    \includegraphics[width=8.5cm]{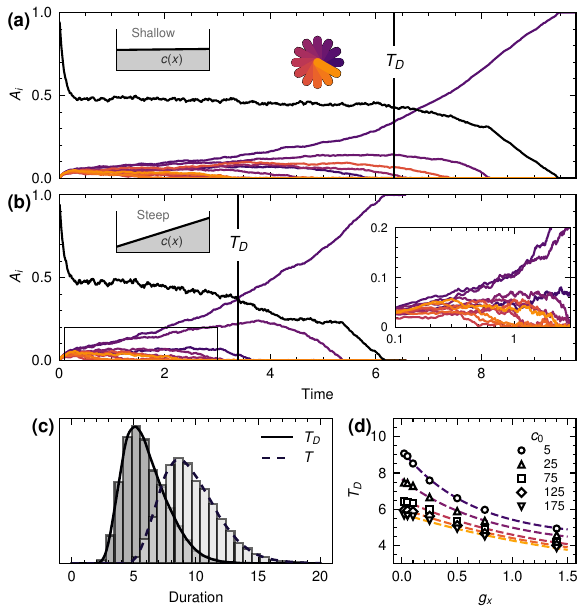}
    \caption{
    \textbf{Actin and pseudopods dynamics during decision-making.}
    \textbf{(a-b)} Sample trajectories of F-actin levels of each of the $n=12$ candidate directions of the cell, shown in the circular diagram in (a), in a linear concentration profile Eq.~(\ref{eq:concentration-profile}) in a shallow (a:~$g_x{=}0.01$, $c_0{=}50$) and a steeper (b:~$g_x{=}1$, $c_0{=}150$) gradient environment, respectively.
    The inset in (b) shows a closeup of the initial dynamics of the competing pseudopods on log time.
    The proportion of G-actin is shown as a black line.
    The decision time $T_D$ is also marked with a vertical line set by Eq.~(\ref{eq:decision_time}).
    \textbf{(c)} Duration distribution of the events $T$ and the decision times $T_D$. 
    \textbf{(d)}~Decision time as a function of the chemoattractant gradient for different noise levels set by the concentration value.
    }
    
    \label{fig:splitting-event-metrics}
\end{figure}

\paragraph{Effect of chemoattractant on actin dynamics.}
We evaluate the system at different concentration profiles and observe the difference in the actin dynamics during the competition event (see Fig.~\ref{fig:splitting-event-metrics}).
The trajectories quickly start by equally recruiting monomers regardless of direction, which halves the pool of G-actin $A_u$.
As shown in Fig.~\ref{fig:splitting-event-metrics}b, the F-actin levels fluctuate until some candidates are suppressed while others remain, forming pseudopods.
This process is quicker when the gradient is stronger.
The resulting distributions of decision time and total duration for many combinations of gradients and concentration values (Fig.~\ref{fig:splitting-event-metrics}c) indicate that decision time only accounts for half the duration of the event.
This demonstrates that, despite cell movement and decision-making being coupled, the decision occurs before the final pseudopod grows and collapses all other candidates, directing the entire cell body toward the chosen direction.
Interestingly, when studying the effect of the environment in the decision time of the cell, we observe an exponential decaying dependency with the gradient strength (Fig.~\ref{fig:splitting-event-metrics}d).
Hence, the cell reacts faster in environments where the gradient (signal) information is stronger.
Similarly, a larger concentration background level $c_0$ (noise) also causes the cell to decrease its decision time.

\begin{figure*}[tbh]
    \centering
    \includegraphics[width=14.0cm]{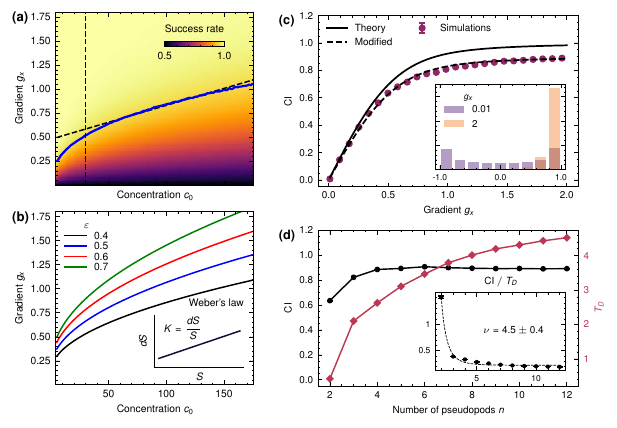}
    \caption{
    \textbf{Weber-like law and speed-accuracy tradeoff.}
    \textbf{(a)}~A heatmap of the success rate at different values of gradient and background concentration.
    Each rate value (square) is calculated by averaging $10^5$ independent splitting events with randomly oriented cells and considering it a success if the final movement of the cell has led it to a higher concentration, i.e., $c(x_T) > c(x_0)$.
    The blue line indicates the minimum gradient at which the accuracy of choosing the direction up the gradient reaches 0.95, the threshold rate at which we consider the cell to be making the correct decision unambiguously, as the perfect rate is subject to numerical fluctuations.
    The vertical dashed line indicates the threshold at which we consider the linear regime to begin ($c(x) > 30$).
    A linear fit is shown in a dashed line on top of the minimum gradient for the linear region.
    \textbf{(b)}~The minimum gradient line changes for different actin exchange parameters $\varepsilon$.
    The inset showcases what is commonly understood as Weber's law, a scalar ratio between the perceived change in stimulus ($dS$) and stimulus value $S$.
    \textbf{(c)}~ Accuracy of aligning the cell body with the gradient, given by the chemotactic index (CI).
     The solid line showcases the optimal CI of a perfect absorbing cell set by Eq.~(\ref{eq:optimal-similarity}), while the dashed line adjusts it by a factor $0.9$.
     The inset exemplifies the resulting distributions of the alignment of the cell at low and high gradients, respectively.
     \textbf{(d)}~CI (black) and mean decision time $T_D$ (red) as a function of the number of candidate pseudopods at the highest gradient ($g_x=2$).
     The inset shows the rate of alignment, fitted to the power law $n^{-\nu}$.
    }
    \label{fig:weber_law_speed_tradeoff}
\end{figure*}

\paragraph{Emergence of Weber-like law.}
When examining cell decision success rate in response to chemoattractant gradients, we find the minimal signal strength required for a consistent movement up the gradient scales with noise strength due to measurement fluctuations (Fig.~\ref{fig:weber_law_speed_tradeoff}a).
This phenomenon echoes Weber's law of just noticeable differences, which describes a linear relationship between signal strength and mean value.
Notably, here, this property emerges without relying on logarithmic transformations of the signal, commonly used in chemotaxis studies~\cite{alonso_learning_2024, clausznitzer_predicting_2014}.
Instead, the difference between concentrations, e.g., at the pseudopod tip and at its original position at the focal adhesion point, suffices to produce Weber law-like behavior, implying that cells may use a simple, local comparison mechanism to make directional decisions.
However, our model results in a square-root dependency on concentration, approximating linearity when noise contributions become significant (see Fig.~\ref{fig:weber_law_speed_tradeoff}a).
The square-root dependency indicates that the cell success rate scales with the signal-to-noise ratio (SNR) instead, which is defined as
\begin{equation}
    \text{SNR}=\dfrac{g_x^2 }{c(x_0)}.
\end{equation}
This can be easily understood in our model, Eq.(\ref{eq:actin-dynamics}), as our cells make finite-difference estimates of concentrations across pseudopods, effectively measuring gradients, while the noise Eq.(\ref{eq:actin-noise}) scales as the square root of the chemoattractant concentration.
This result also matches previous experimental observations \cite{tweedy_distinct_2013} and is consistent with Weber's law observed in the chemotaxis of some cell types~\cite{adler_fold-change_2018, goentoro_evidence_2009}.

Interestingly, adjusting $\varepsilon$, which determines how fast the cell commits to the winner pseudopod, modifies the slope of the dependency while preserving its linearity (Fig.~\ref{fig:weber_law_speed_tradeoff}b).
This showcases robustness in the behavior of the cell, which, despite changing the decision accuracy and speed, conserves the overall dependencies on the environment information.

\paragraph{Pseudopod competition leads to indirect gradient sensing.}
A commonly used measurable observable for chemotactic performance is the \textit{chemotactic index}, which was previously calculated using the physical limits of sensing by \citet{endres_accuracy_2008}.
The probability of estimating the gradient of a concentration by a perfectly ligand-absorbing cell (to avoid the noise from rebinding) is
\begin{equation}\label{eq:estimation_limit}
    P(\hat{g}_x, \hat{g}_y) = \dfrac{1}{2\pi\sigma_g^2} \exp{\left[{\frac{-(\hat{g}_x - g_x)^2 - (\hat{g}_y - g_y)^2}{2\sigma_g^2}} \right]},
\end{equation}
where $(\hat{g}_x, \hat{g}_y)$ is the estimated gradient, and the real one is given by $\nabla c = (g_x, g_y)$, with the uncertainty of the measurement being
\begin{equation*}\label{eq:uncertainity-gradient-estimation-limits}
    \sigma_g^2 = \dfrac{c_0}{12 \pi D T}.
\end{equation*}
where $T$ and $D$ are the relative time the cell takes to measure the gradient and the unitless diffusion constant of the chemoattractant, respectively.
From Eq.~(\ref{eq:estimation_limit}) and assuming a linear gradient on $x$ such that $\nabla c~=~(g_x, 0) $, we obtain that the expected cosine similarity of the gradient and direction of movement.
This is the chemotactic index given by
\begin{equation}
    \text{CI} = \langle \cos{\left( \theta \right)} \rangle = \sqrt{\dfrac{\pi z}{2}} e^{-z} \left[ I_0(z) + I_1(z) \right],
    \label{eq:optimal-similarity}
\end{equation}
where $ z = 3\pi k $ SNR, $\theta$ is the orientation of the cell, and $I_{0(1)}$ are first (second)-order modified Bessel functions.
The combination $k=DT$ can be thought of as a single fitting parameter.
During the measuring time, it is assumed that the cell processes the measurements by averaging their positional information before making a decision.
Instead, Fig.~\ref{fig:weber_law_speed_tradeoff}c shows that this processing emerges from the competition between pseudopods until one candidate is chosen and the cell aligns itself with the estimated gradient.
However, the resulting CI from numerical simulations saturates at a lower value due to the similarity in signal between neighboring candidates and the saturation of the receptor's signaling, set by Eq.(\ref{eq:polymerization-rate}).
Agreement with the theory is only obtained when multiplying Eq.(\ref{eq:optimal-similarity}) by a $0.9$ factor.
Interestingly, this was also done in \citet{endres_accuracy_2008} when comparing to data.

\paragraph{Pseudopods competition reveals speed-accuracy trade-off.}
Due to the pseudopods' fixed orientation, the larger the number of candidates, the more likely a pseudopod will align perfectly with the gradient direction.
As shown in Fig.~\ref{fig:weber_law_speed_tradeoff}c, the closeness of the candidates, together with the low signal or strong fluctuations of the measurements, saturate the decision.
Thus, we compare the resulting alignment at the highest signaling strength for an increasing amount of candidates (Fig.~\ref{fig:weber_law_speed_tradeoff}d) and notice that the alignment increases with $n$ up to a saturation point at $n=6$.
Curiously, when estimating the average decision time $\langle T_D \rangle$, we observe a monotonic increase as well, without reaching a saturation state, pointing towards a diminishing return in the number of candidates in terms of efficiency. 
Furthermore, by evaluating the rate of alignment (Fig.~\ref{fig:weber_law_speed_tradeoff}d), we infer a power-law-like behavior $\text{CI}\mathbin{/}T_D\approx n^{-\nu}$, with $\nu=4.5$, which clearly indicates that the lower the number of candidates, the lower the decision time, and the more efficient the decision making (despite a decrease in absolute accuracy).

\begin{figure*}[tbhp]
    \centering
    \includegraphics[width=16cm]{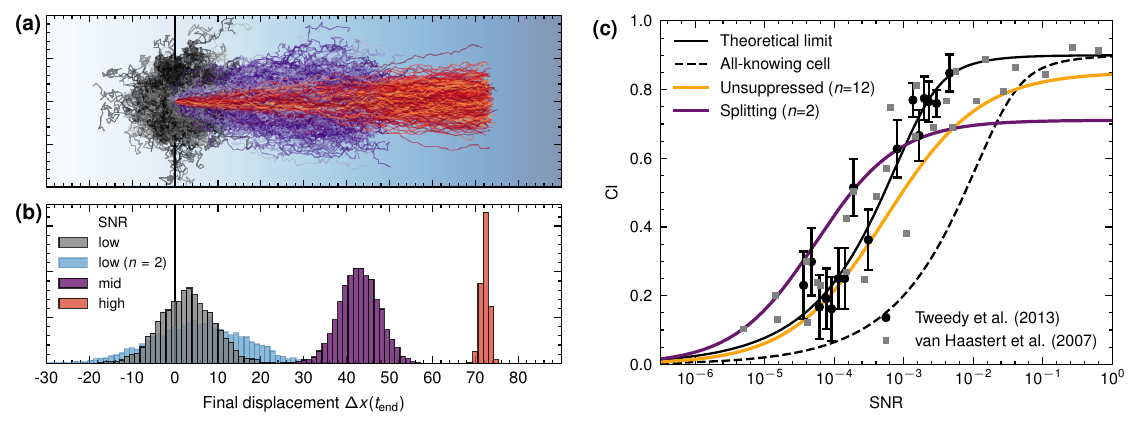}
    \caption{
    \textbf{Chemotaxis performance depends on SNR.}
    \textbf{(a)}~Sample of $3\,10^2$ chemotaxis trajectories of our model, composed of 30 \textsl{splitting events}, at different levels of SNRs.
    The vertical black line indicates the initial positions of the cell.
    Since SNR changes along a trajectory, low here contains SNR $\in~[10^{-5.874}, 10^{-5.875}]$, mid contains $[10^{-3.26},10^{-3.29}]$ and high $[10^{-1.87}, 10^{-2.06}]$.
    \textbf{(b)}~Final displacement distribution ($N=7\, 10^3$) of the trajectories in (a).
    A distribution for the splitting configuration with $n=2$ at low SNR is also included in blue.
    \textbf{(c)}~Chemotactic index (CI) as a function of SNR, plotted on a $\log_{10}$ scale.
    The fundamental physical limit for a static spherical ligand absorber is given by Eq.~(\ref{eq:optimal-similarity}) \cite{endres_accuracy_2008}, fitted to the experimental data from \cite{tweedy_distinct_2013, van_haastert_biased_2007}.
    Additionally, the performance of an \textit{all-knowing} cell that estimates the gradient based on ($n=12$) sensors uniformly distributed across its body is shown in a dashed line (see SI for details).
    In yellow is the resulting CI of $10^4$ independent trajectories, from which at each timestep the CI is calculated using Eq.(\ref{eq:ci}) and binned according to the SNR at the start of the event. 
    Similarly, the purple line shows the results for a cell with only activated $\mathcal{P}_2$ and $\mathcal{P}_{10}$.
    For clarity, the lines are fitted to a logistic function using the numerical simulations binned by SNR.
    }
    \label{fig:chemotaxis-results}
\end{figure*}

\paragraph{Chemotaxis trajectories depend on SNR.}
The dynamics of the \textsl{splitting event} move the cell body towards its chosen direction.
When completed, the winning pseudopod becomes the cell body in a new location with a new orientation, ready to start another event.
Consequently, in our model, chemotaxis emerges as a sequence of consecutive pseudopod-splitting events.
Similar to the decision success rate in Fig.~\ref{fig:weber_law_speed_tradeoff}a, by extending pseudopod splitting to many events, the resulting ensemble displacement is strongly affected by the SNR (Fig.~\ref{fig:chemotaxis-results}a).
The spread of the cell during the trajectory is highly correlated with the accuracy of each individual decision.
At low SNR, the resulting distribution is almost that of a random walk, whereas, at high values, the cells show strong persistence in moving up the gradient.

In Fig.~\ref{fig:chemotaxis-results}c, we compare the chemotactic index of our unsuppressed cell, i.e., with $n=12$ possible directions of movement (yellow line), with those from experiments (symbols).
To quantify the alignment of trajectories, we calculate the chemotactic index as the average of the cosine similarity $S_C$ for binned values of SNR.
This can be expressed as  $CI = \langle S_C(t) \rangle_{\text{SNR}}$.
Due to the shape of our cell, $S_C$ is defined as the weighted average of the orientation of each pseudopod (with respect to the uropod), proportional to their length.
Thus, at each timestep, we have
\begin{equation}\label{eq:ci}
    S_C(t) = \ell_u \ \cos{(\theta)} + \sum_i^n \ell_i \ \cos{(\theta + \phi_i)} ,
\end{equation}
where $\theta$ indicates the cell orientation and $\phi_i$ the relative orientation of the candidate $\mathcal{P}_i$ w.r.t. the cell.
Interestingly, at lower SNR, when the gradient is shallow and difficult to infer, the average performance resembles the expected one of a perfect absorbing cell at the fundamental limit of sensing (black solid line).
Note that for wide-ranging attractant concentrations, our cells with only two pseudopods perform significantly better than all-knowing cells, implemented by $12$ sensors, sophisticated inference by least squares fitting, and uncertainty based on the Cramér-Rao bound (see SI for details).
Furthermore, as the signal increases, the model approaches the lower bound of the experimental data, ultimately saturating at $\sim 0.9$ chemotactic index.
There is notably excellent agreement between our model and the data without any explicit fitting.

\paragraph{Pseudopod suppression to enhance chemotactic efficiency.}
Until now, unless explicitly specified, we assumed that the cell has $n{=}12$ evenly spaced distributed pseudopod candidates, sufficient for the cell to accurately choose the right moving direction.
However, we observed that fewer candidates can drastically improve efficiency during decision-making (Fig.~\ref{fig:weber_law_speed_tradeoff}d).
Experimental observations showed that the angle between pseudopods is affected by the chemoattractant gradient shallowness \cite{van_haastert_chemotaxis_2010, bosgraaf_navigation_2009}.
Based on these observations, we suppress all pseudopods candidates except $\mathcal{P}_2$ and $\mathcal{P}_{10}$, which grow at a $\varphi {=}\pm60^\circ$ of the cell movement orientation, and observe their CI as the SNR  increases (Fig.~\ref{fig:chemotaxis-results}c).
The resulting chemotactic index shows that the cell is able to move up the gradient more robustly at small SNRs than when allowed to change directions freely ($n=12$).
This is also seen on the final displacement at Fig.~\ref{fig:chemotaxis-results}b at very low SNR, where this configuration manages to reach further up the gradient.
Nevertheless, as the quality of the signal increases, the cell with the wider range of options performs better, as the two pseudopods likely do not align with the actual gradient. 
This cross-over in the chemotactic index can be seen in the data of cell morphology, where amoeba use pseudopod splitting at small SNR and a broad-front polarization at high SNR \cite{tweedy_distinct_2013}.

In previous work \cite{alonso_learning_2024},  we observed the advantage of gradual orientation updates instead of full turns when optimizing a fully unconstrained spatial policy using deep reinforcement learning (DRL) at the fundamental limits of ligand sensing, confirming that relying on persistence is an effective strategy when navigating shallow gradients.
What can DRL tell us about the ideal number of pseudopods and their orientation?

\begin{figure*}[tbh]
    \centering
    \includegraphics[width=18cm]{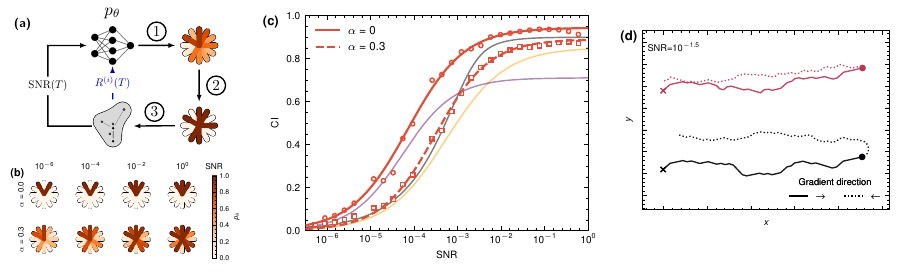}
    \caption{
    \textbf{Optimal pseudopod suppression strategy.}
    \textbf{(a)}~Schematics of the DRL training process, where (1) the policy outputs activation probabilities of each candidate, (2) a sampling occurs to suppress certain directions, and (3) a splitting event is simulated, resulting on a new outcome which we use to optimize the weights of the policy, and a new SNR that will be used as input for the next step.
    \textbf{(b)}~The cell diagrams show the resulting activation probability $p_\theta$ of the 12 candidates at different SNR for both $\alpha=0$ and $\alpha=0.3$.
    On the schematics, the cell direction movement is up.
    \textbf{(c)}~Evaluation of the optimal policy $p_\theta$ using PPO with $\gamma=0.2$, based on how well pseudopods align with the gradient compared to the results from Fig.~\ref{fig:chemotaxis-results}c.
    The resulting points are fitted with a logistic function and are the result of $10^4$ independent simulated trajectories, from which, at each time step, the CI is calculated and binned by SNR.
    Both the policy strained on static profile concentrations ($\alpha=0$), and the one trained with a high rate of changing gradient direction ($\alpha=0.3$) are shown despite being evaluated on a static gradient.
    \textbf{(d)}~Trajectories of the suppression policy trained in a static gradient (black) and in a dynamic gradient with a high rate of switching (red).
    The initial position of the cell is marked with a cross, while the point of the trajectory at which the gradient sign is changed is indicated by a dot.
    }
    \label{fig:suppression_results}
\end{figure*}

\paragraph{Optimal pseudopod suppression policy.}
Having observed a clear advantage in suppressing possible directions, we explore the possibility of an optimal configuration in which the cell may have learned to exploit suppressing certain directions throughout the course of evolution to enhance its chemotactic performance.
Here, we optimize a mapping function $p_\theta:~\mathbb{R} \to [0, 1]^n$ that indicates the probability of not suppressing candidate $i$ by setting the signal strength to either suppressed ($\rho_0=0$) or active ($\rho_0=1$).
Due to the number of possible combinations of pseudopod states, classical optimization techniques are computationally unfeasible.
Therefore, we rely on modern reinforcement learning approaches, specifically, proximal policy optimization~(PPO)~\cite{schulman_proximal_2017}, to locate the optimal suppression policy, here constructed as an artificial feed-forward neural network, given the environment's state, set by the SNR at the cell location at the start of the event (see Fig.~\ref{fig:suppression_results}a and Methods for further details).

We define the reward function after each decision step during the chemotactic trajectory of the cell $(i)$ as 
\begin{equation}\label{eq:reward}
    R^{(i)}(x_T, T) = \cos(\theta_T) + \gamma \left( \dfrac{t_{max}-T}{t_{max}} \right),
\end{equation}
where $t_{max}$ is the maximum possible time for the cell to make a decision before a random one is selected, and $\gamma$ is a time penalty to favor the configurations that lead to faster events.
Subsequently, we optimize the weights on the network by maximizing the cumulative reward $\hat{R}^{(i)}$ during a trajectory, which we specify as the concatenation of $30$ splitting events.

The resulting configurations further confirm the important role of persistence in accurate chemotaxis, as it enables the cell to place its receptors at the leading edge, and by relying on consecutive turns, the cell further aligns itself with the gradient.
Despite exploring a wide range of SNR, only two final configurations emerge that surpass both of our previous proposed ones (Fig.~\ref{fig:suppression_results}b).
At low SNR, two candidate pseudopods, closer than those we constructed from experimental observations, prove optimal, further demonstrating that relying on small changes reduces the likelihood of sudden errors.
Given that the limited information the cell gains from the environment is not enough to distinguish forward and side directions, the found policy optimizes for time, which results in a configuration with a minimal number of candidates ($n=2$), echoing our results on alignment efficiency from Fig.~\ref{fig:weber_law_speed_tradeoff}d.
However, as the information from the environment increases, the subtle difference between neighboring candidates becomes apparent.
The resulting policy at high SNR activates the forward pseudopod candidate while still minimizing the number of candidates to yield better alignment during long trajectories, where the concentration profile remains static.
When comparing the CI of the trajectories as a function of the SNR, we obtain an envelope curve, resulting in an upper limit to previous performances (Fig.~\ref{fig:suppression_results}c).

The ability of a cell to react and adapt to sudden changes in the environment is another criterion of successful chemotaxis.
Specifically, \citet{aquino_memory_2014} investigated how microorganisms respond to abrupt changes in the chemoattractant direction, demonstrating that cells can adapt when the direction is suddenly reversed, at least in steep gradients.
We introduce the possibility, set by $\alpha$, for the gradient direction to randomly change during a chemotaxis trajectory, which we include while training the policy.
Given that our previous policy exploited the persistence of movement in a static profile, we examine the performance of both the old policy ($\alpha=0$) and the new one ($\alpha=0.3$) in enabling the cell to adjust its trajectory in response to a sudden change in the gradient sign.
Interestingly, when $\alpha=0.3$, the resulting pseudopod configurations are more spread, showing the possibility of activating the rear pseudopod candidates at low SNR and consistently activating them at high SNR, where the frontal candidates are likely to win when the difference in signal is notable (Fig.~\ref{fig:suppression_results}b).
Observing example trajectories, a forward-facing configuration reacts slower to changes in the gradient, performing U-turns to realign with the source direction on both low and high SNR scenarios (Fig.~\ref{fig:suppression_results}d).
This contrasts with the adaptive policy, which can immediately reverse direction when the change occurs at high SNR, reproducing previously observed behaviors \cite{aquino_memory_2014} and the cells' ability to form {\it de novo} pseudopods \cite{van_haastert_chemotaxis_2010}.
Despite the benefit of a better reaction time to sudden changes in the environment information, when evaluated on a static concentration profile as those more common on cellular chemotaxis, we recover a similar performance as the one set by the limits of sensing (Fig.~\ref{fig:suppression_results}c).
Thus, cells lose the advantage persistence yields for overcoming those limits.
Combining DRL, an effective optimization technique for non-trivial problems, with the interpretability of classical systems biology, we obtain insights into how different evolutionary traits may have altered the navigation strategies of motile, chemotactic cells.

\section{Discussion}

We introduced a minimal model for cellular decision-making based on the competition between pseudopods with stimulus-dependent growth.
Instead of relying on an all-knowing cell that evaluates the measurements of the chemical gradient to decide its orientation of motion, we propose that pseudopod dynamics, driven by the stochastic processes of actin polymerization, simple G-actin conservation, and mutual inhibition, are responsible for the cell's emergent decision.
Hence, no direct spatial gradient sensing is required.
Despite its simplicity, we have shown that our model agrees with the theoretical limit while providing insights into the time costs of moving up the chemoattractant gradient.
Furthermore, our model captures the key features of pseudopod dynamics as observed in experiments, such as the emergence of multiple pseudopods on shallow low-SNR gradients and the scaling of the signal with noise to move up the gradient unequivocally, reminiscent of Weber's law.

When concatenating consecutive decision steps, i.e., splitting events, we modeled chemotaxis trajectories capable of reproducing experimental data, where our results showed a characteristic dependence of the chemotactic index on the SNR, previously postulated in \citet{endres_accuracy_2008}.
We then extended the model by incorporating a learnable suppression mechanism, allowing us to explore how cells might have optimized their polarization for efficient chemotaxis, particularly at low SNR, where gradient information is especially limited.
By employing deep reinforcement learning, we found a mapping between the SNR and pseudopod suppression.
Interestingly, the learned suppression policy converges towards a behavior where the cell preferentially suppresses pseudopods at small angles, leading to a more focused forward distribution of directions of motion.
These findings align with experimental observation in shallow gradients, whereby cells exhibit a fixed angular spread of forward-facing pseudopods \cite{van_haastert_chemotaxis_2010}.

Spread-out cell protrusions are optimal for instantaneous sensing, as they increase the spatial information of the environment around the cell \cite{bodor_cell_2020, nakamura_gradient_2024}.
Our results suggest that cell polarization promotes forward-facing pseudopods, which, while clearly suboptimal for an instant decision, prove advantageous during chemotaxis.
By leveraging persistence, which bypasses the need for a memory of prior information, the cell is capable of maintaining a consistent direction of movement, thereby improving its ability to navigate gradients over longer trajectories.
Thus, our findings highlight the tradeoff between instantaneous, accurate sensing and overall chemotactic performance.

The model proposed here does present certain limitations, as assumptions were made to facilitate interpretability and simplification.
Thus, future research may want to expand on this work by increasing its realism and complexity.
Notably, we have ignored the physical effects of the moving cell, e.g., the "windshield effect" of a ligand-absorbing cell \cite{endres_accuracy_2008, tweedy_distinct_2013} and the mechanical properties of the cell membrane as the cell crawls on the substrate, as well as molecular details of the spatial positioning of neighboring pseudopods with mutual inhibition.
Such positioning and regulation could be achieved by simple reaction-diffusion mechanisms of activators and inhibitors \cite{tweedy_distinct_2013}.
Furthermore, our simplification of concentration sensing in terms of stimulus-dependent growth of pseudopods could be investigated in more detail to capture the effects of ligand binding and unbinding times or ligand-induced receptor internalization.
Further experimental validation of our model may include tracking the dynamics of pseudopods in testable decision-making scenarios according to the SNR.
We hope further research into the mechanistic foundation of decision-making may yield novel insight into cellular behavior.

Amoeboid cell migration based on pseudopods and other cell protrusions is not unique to \textit{D. discoideum} but also occurs in neutrophils \cite{andrew_chemotaxis_2007} and even in the spermatozoa of \textit{C. elegans}\cite{nelson_caenorhabditis_1982}.
Notably, the latter achieves motility without actin, emphasizing shape and behavior as fundamentally important and that the biochemical details might be secondary \cite{fritz-laylin_evolution_2020}.
Instead, we suggest that amoeboid shape and behavior are evolutionarily conserved traits, providing advantages in fast and accurate chemotaxis.
Even the syncytial plasmodia of the slime mold \textit{Physarum polycephalum} forages and grows as a macroscopic network \cite{alim_fluid_2018}.
Not unlike our proposed stimulus-driven pseudopod extension in Dicty, nutrient uptake on one end of the network drives extensions in this favorable direction, leading to retraction at the rear.
Whether our navigation strategies are also relevant to ciliated micro-organisms \cite{wan_origins_2021} or group chemotaxis~\cite{gonzalez_collective_2023} are fascinating open questions.
However, our results clearly go beyond chemotaxis in cells.
With an increasing interest in developing microscopic artificial agents, a theoretical framework for decision-making in difficult-to-navigate environments is crucial.
Amoeba-inspired limbless robotics may benefit from our robust strategy designs without requiring extensive hard-wired sensor-driven feedback mechanisms \cite{umedachi_fluid-filled_2012, deng_amoeboid_2023, rode_information_2024}.
We showed that by suppressing the strength of candidate directions, the cell can use persistence to effectively navigate up a gradient by focusing on creating fewer forward-facing protrusions.

In conclusion, our work proposes a new understanding of the fundamental principles governing cell decision-making and their implications for chemotaxis in complex environments.
We showed that pseudopod splitting leads to highly effective chemotaxis without a need for direct spatial sensing and memory, demonstrating aspects of mechanical intelligence. The paper also highlights the potential of reinforcement learning as a powerful tool for studying and understanding the intricate interplay between cellular mechanics, sensing, and behavior without relying on black-box decision policies that obscure the internal cellular mechanism.
Applications in robotics are apparent.

\section{Methods}

\paragraph{Numerical simulations.}
Actin dynamics are integrated using the Euler-Maruyama integration scheme, converging to the Ito solution.
Discrete-time steps are set to $\Delta t=0.1~s$ during the simulations unless explicitly stated.
Thousands of realizations are carried out for each numerical result by massively parallelizing the simulations using GPUs (see Code availability).

\paragraph{Non-dimensionalization of the dynamics.}
To simplify the equations, we non-dimensionalize the system by scaling all lengths relative to cell size $a$.
Similarly, time is defined relative to $\tau_m$, a characteristic timescale for the linear filter in Eq.~(\ref{eq:length-linear-response}), which relates to the mechanical properties of the membrane.
Accordingly, the gradient and concentration levels are rescaled by length scale $a$, resulting in unitless concentration profile  Eq.~(\ref{eq:concentration-profile}).
Hence, length scale $a$ changes the concentration profile while maintaining the same cell dimensions.
The cell dynamics are described in the cell's frame of reference.

\paragraph{Suppression policy architecture.}
We model the suppression policy $\rho_\theta$ as a relatively small artificial neural network whose input is the logarithm in base 10 of the SNR and whose output is a vector of $(n, 2)$ values between 0 and 1 as the logits for the probability of activating or suppressing each pseudopod $\boldsymbol{\rho_0}$.
The final state is sampled from a categorical distribution.
The network is a multi-layer perceptor (MLP) of 4 layers of 128 neurons each, with $\tanh$ activation functions between them.
Since the policy also predicts the expected value $V$, we use an MLP with 4 layers and 128 neurons.

\paragraph{Optimizing the suppression policy.}
The algorithm used here is a modified version of the proximal policy optimization (PPO) algorithm~\cite{schulman_proximal_2017} implemented in~\citet{alonso_learning_2024}.
PPO is an on-policy optimization technique that iteratively improves its policy $p_\omega$ by collecting information from simulations between optimization steps.
The results of these simulations are then used to perform stochastic gradient descent on the policy parameters $\omega$.
Simply, the algorithm maximizes the following clipped surrogate loss defined as:

\begin{equation*}
\begin{split}
L^{CLIP}(\omega) = \mathbb{E} \Bigg[ \min\Bigg( & \frac{p_{\omega_n}(\boldsymbol{\rho_0}|z)}{p_{\omega_p}(\boldsymbol{\rho_0}|z)} \cdot A_t, \\
& \text{clip} \left( \frac{p_{\omega_n}(\boldsymbol{\rho_0}|z)}{p_{\omega_p}(\boldsymbol{\rho_0}|z)}, 1 - \epsilon, 1 + \epsilon \right) \cdot A_t \Bigg) \Bigg],
\end{split}
\end{equation*}
where $ \omega_n $ represents the updated parameters of the policy, while $\omega_p$ indicates the previous policy parameters.
The term $p_{\omega}(\boldsymbol{\rho_0}|z)$ is the probability of output $\boldsymbol{\rho_0}$ given the SNR, here indicated by $z$, under the new policy.
The advantage function $A_t$ quantifies the relative benefit of taking a particular action at a given state compared to the average action value.
Notably, the clipping parameters $\epsilon$ control the size of the trust region, ensuring that new updates do not deviate significantly from the previous policy, which leads to more stable optimizations.
In practice, the clipping is performed by defining upper and lower bounds on the allowed change in ratio between consecutive policies that contribute to the loss.

To promote exploration, especially given the discrete nature of our actions (active or suppressed candidate), we include an entropy term on the loss function as a regularization term set by

\begin{equation*}
    H = -\dfrac{1}{n}\sum_i^n p^{(i)}_\omega(z) \log(p^{(i)}_\omega(z))
\end{equation*}
which encourages the policy to maintain a certain degree of stochasticity, preventing collapses and premature convergence.
Here, $n$ is the number of actions, which is set as the number of candidate pseudopods.

\paragraph{Experimental data.} 
The experimental frames shown in Fig.~\ref{fig:diagram-splitting-event} were provided by Robert Insall \cite{andrew_chemotaxis_2007} and taken from  \cite{sorrentino_structural_2021} with permission.
The data points of the chemotactic index used in Fig.~\ref{fig:chemotaxis-results} were obtained from previous studies by \citet{tweedy_distinct_2013, van_haastert_biased_2007}.

\subsection*{Acknowledgments}
We thank Antonio Matas Gil,  Martina Oliver Huidobro, and Silja B. Låstad for their stimulating discussions and valuable feedback on this work.
This project has received funding from the Novo Nordisk Foundation Grant Agreement NNF20OC0062047.

\subsection*{Code availability}
All the numerical simulations and deep reinforcement learning implementations are available and can be found at \url{https://github.com/Endres-group/psxc-research}

\subsection*{Declaration of interests}
The authors declare no competing interests.

\bibliography{references}{}

\end{document}